\documentclass[aps,prl,twocolumn,a4paper,showkeys,showpacs,footinbib,superscriptaddress]{revtex4-1}

\usepackage{amsmath,amssymb,bm,amsthm}
\usepackage{subfigure}
\usepackage[dvipdfmx]{graphicx}
\usepackage{dcolumn}
\usepackage{bm}
\usepackage[mathlines]{lineno}
\usepackage{booktabs}
\usepackage{color}
\newcounter{one}
\usepackage{url}

\usepackage{textcase}

\usepackage{newtxtext,newtxmath}

\usepackage{colortbl}
\usepackage{tabularx}
\usepackage{verbatim}
\usepackage{multirow}

\usepackage[T1]{fontenc} 
\usepackage[utf8]{inputenc}
\usepackage{amsfonts}
\usepackage{array}

\usepackage{enumitem}
\usepackage[usenames,svgnames]{xcolor}
\usepackage{natbib}
\usepackage[dvipdfmx,colorlinks=true,linkcolor=blue,citecolor=blue,breaklinks=true]{hyperref}

\usepackage{amsmath,amssymb,bm,amsthm}

\def\beq{\begin{equation}}
\def\eeq{\end{equation}}
\def\nbeq{\begin{equation*}}
\def\neeq{\end{equation*}}
\def\beqa{\begin{eqnarray}}%
\def\eeqa{\end{eqnarray}}%

\def\<{\langle}
\def\>{\rangle}

\def\Tr{{\rm Tr}}

\newcommand{\sectionprl}[1]{{\par\it #1.---}}

\begin{document}
\title{Strong Eigenstate Thermalization within a Generalized Shell in Noninteracting Integrable Systems}

\author{Takashi Ishii}
\email{ishii3@iis.u-tokyo.ac.jp}
\affiliation{
Department of Physics, Graduate School of Science,
University of Tokyo, Kashiwa 277-8574, Japan
}
\author{Takashi Mori}
\email{takashi.mori.fh@riken.jp}
\affiliation{
RIKEN Center for Emergent Matter Science (CEMS), Wako 351-0198, Japan
}

\begin{abstract}
Integrable systems do not obey the strong eigenstate thermalization hypothesis (ETH), which has been proposed as a mechanism of thermalization in isolated quantum systems.
It has been suggested that an integrable system reaches a steady state described by a generalized Gibbs ensemble (GGE) instead of thermal equilibrium.
We prove that a generalized version of the strong ETH holds for noninteracting integrable systems with translation invariance.
Our generalized ETH states that any pair of energy eigenstates with similar values of local conserved quantities looks similar with respect to local observables, such as local correlations.
This result tells us that an integrable system relaxes to a GGE for any initial state that has subextensive fluctuations of macroscopic local conserved quantities.
Contrary to the previous derivations of the GGE, it is not necessary to assume the cluster decomposition property for an initial state.
\end{abstract}
\maketitle

\sectionprl{Introduction}
Out-of-equilibrium dynamics of isolated quantum systems and their steady states have been explored recently~\cite{DAlessio_2016,Eisert_review2015,mori2018review}.
Various experiments~\cite{Trotzky2012, Kaufman2016, Neill2016, Clos2016} as well as numerical calculations~\cite{Saito1996,Rigol2008, Jin2010} have revealed that nonintegrable systems thermalize under unitary time evolution.
As a possible mechanism of thermalization, the eigenstate thermalization hypothesis (ETH) has been studied~\cite{Deutsch1991, Srednicki1994, Rigol2008, Biroli2010, Steinigeweg2013, Kim2014, Beugeling2014, Shiraishi-Mori2017, Mori-Shiraishi2017}.
The ETH in the strong (weak) sense, or namely ``the strong (weak) ETH'', states that all (almost all) energy eigenstates have thermal properties when we look at local observables.
It has been recognized that the strong ETH ensures thermalization, while the weak ETH does not because the initial state may have an important weight on nonthermal energy eigenstates~\cite{Biroli2010}.
Indeed, the weak ETH can be proved for generic translationally invariant systems including integrable systems~\cite{Biroli2010,Mori2016,Iyoda2017}, although it is known that integrable systems generally do not thermalize~\cite{Rigol2007,Biroli2010,Cassidy2011,Essler2016}
~\footnote{We note that some studies have shown~\cite{Berman2004,Santos2012} that a kind of relaxation that is defined in terms of certain statistical indexes may not be qualitatively affected by integrability.}. 
Meanwhile, the strong ETH has been numerically verified in nonintegrable models~\cite{Beugeling2014,Kim2014}.

For integrable systems, it is suggested that the steady state is given by the so-called generalized Gibbs ensemble (GGE)~\cite{Rigol2007}, which is constructed by using a set of local and quasi-local conserved quantities of the system. 
Numerical studies on specific integrable models support the validity of the GGE~\cite{Rigol2007, Cassidy2011}.
By analytically calculating the time evolution of local observables, the validity of GGE is even proved for noninteracting integrable systems with translation invariance when the initial state satisfies some certain properties. 
More precisely, proofs were given for Gaussian initial states~\cite{Barthel2008,Sotiriadis2009,Calabrese2011,Calabrese2012_1,Calabrese2012_2}, followed by extensions to initial states that satisfy the cluster decomposition property~\cite{Cramer2008,Sotiriadis2014,Gluza2016}; see also Ref.~\cite{Bastianello2017} for continuous models.
Though the present work focuses on noninteracting integrable systems, it should be noted that the validity of the GGE has also been investigated for interacting integrable systems~\cite{Pozsgay2013, Fagotti2013, Fagotti-Essler2013, Goldstein2014, Pozsgay2014, Wouters2014, Pozsgay2014_correlations, Brockmann2014, Mestyan2015, Ilievski2015}, which cannot be mapped to free particles but exactly solvable via the Bethe ansatz method.

Here, a set of questions naturally arises. 
Can one construct a generalized version of the ETH as a mechanism that explains the relaxation to a GGE, just as the standard ETH explained thermalization in nonintegrable systems?
If so, can we remove the assumption of the cluster decomposition property for the initial state in deriving the relaxation to a GGE?
The removal of this assumption is important in considering a spin system that can be mapped to a quadratic fermion Hamiltonian (e.g., the transverse-field Ising model) because it is not obvious whether a physically realistic initial state, which satisfies the cluster decomposition property \textit{with respect to spin operators}, also satisfies it \textit{with respect to fermion operators}~\cite{Murthy_arXiv2018}; indeed, there are cases where a nonlocal transformation reveals nontrivial correlations~\cite{Nijs1989,Hida1992,Hatsugai1992}.

In Ref.~\cite{Cassidy2011}, a generalization of the ETH has been proposed. 
Their generalized ETH has been numerically verified~\cite{Cassidy2011} and also proved for various local operators in the translationally invariant transverse-field Ising model~\cite{Vidmar-Rigol2016}, but \textit{only in the weak sense}.
It has not been clarified yet whether it is valid in the strong sense.
Although the concept of the generalized ETH helps us to understand the validity of the GGE~\cite{Cassidy2011, Vidmar-Rigol2016}, the weak generalized ETH does not ensure in itself the relaxation to a GGE in an integrable system.
It is therefore desirable to formulate the generalized ETH that is valid in the strong sense.

In the present paper, by constructing a generalized shell that is specified by a set of macroscopic conserved quantities, we reformulate the generalized ETH and analytically prove that our generalized ETH proposed is valid \textit{in the strong sense} in integrable models of the quadratic form with translation invariance.
It is shown that our strong generalized ETH ensures the relaxation to a GGE for initial states that have subextensive fluctuations of macroscopic local conserved quantities~\footnote{The condition of subextensive fluctuations of macroscopic local conserved quantities is much weaker than the condition of the cluster decomposition property. 
The latter implies the former, but the former does not imply the latter. 
}.
We manage to remove the assumption of the cluster decomposition property here, and thus our result is beyond the previous rigorous results~\cite{Cramer2008,Sotiriadis2014,Gluza2016}. 
In Table~\ref{table:a} we show for help of understanding a comparison between the strong ETH and our strong generalized ETH. 
\begin{table*}
  \begin{tabular}{|c||c|c|c|}\hline
     & Hilbert subspace & steady state & validity \\\hline
    strong ETH~\cite{Beugeling2014,Kim2014,Shiraishi-Mori2017, Mori-Shiraishi2017,Rigol2008, Biroli2010, Steinigeweg2013, Beugeling2014} & energy shell & Gibbs ensemble & \begin{tabular}{c}nonintegrable: valid but with counterexamples.\\integrable: invalid.\end{tabular} \\\hline
    \begin{tabular}{c}strong generalized \\ETH (present study)\end{tabular} & \begin{tabular}{c}shell defined by many \\macroscopic conserved quantities\end{tabular} & \begin{tabular}{c}generalized \\Gibbs ensemble\end{tabular} & \begin{tabular}{c}translationally invariant \\noninteracting integrable: valid.\end{tabular}  \\\hline
  \end{tabular}
  \caption{A comparison between the strong ETH~\cite{Beugeling2014,Kim2014,Shiraishi-Mori2017, Mori-Shiraishi2017,Rigol2008, Biroli2010, Steinigeweg2013, Beugeling2014} and our strong generalized ETH. 
  While the usual strong ETH is discussed for states in the energy shell, in the formulation of our strong generalized ETH we consider a generalized shell, which is defined as a Hilbert subspace specified by a set of macroscopic conserved quantities. 
  The strong ETH and our strong generalized ETH are sufficient conditions for relaxation to the steady state described by the Gibbs ensemble and the generalized Gibbs ensemble, respectively. 
  In the column indicated ``validity,'' we explain the current understanding on the validity of the two concepts. 
  As for the strong ETH, in nonintegrable systems, its validity has been numerically confirmed~\cite{Beugeling2014,Kim2014}, although there exists some counterexamples~\cite{Shiraishi-Mori2017, Mori-Shiraishi2017}. 
  In integrable systems, numerical demonstrations and analytical calculation show that the strong ETH does not hold~\cite{Rigol2008, Biroli2010, Steinigeweg2013, Beugeling2014}. 
  As for our strong generalized ETH, we analytically prove in this paper its validity in translationally invariant noninteracting integrable systems.  
  }
\label{table:a}\end{table*}

\sectionprl{Model and Setup}
\label{sec:model}
We consider a bilinear fermion system described by the translationally invariant Hamiltonian
\beqa
H&&=\sum_{x,y=1}^L \left(c_x^{\dag}A_{x-y}c_y+c_x^{\dag}B_{x-y}c_y^{\dagger}+c_xB^*_{y-x}c_y\right)
\eeqa
under the periodic boundary condition; the analysis is almost unchanged for the anti-periodic boundary condition.
The coefficients $A_{l}$ satisfies $A_{l}=A_{-l}^*$ because $H=H^{\dagger}$.
We assume the locality of the Hamiltonian, i.e., $A_{l}=B_{l}=0$ for $|l|_P>r_H$ with a finite range $r_H>0$, where $|l|_P:=\min\{|l|,L-|l|\}$ denotes the distance $l$ under the periodic boundary conditions.
This form of Hamiltonian includes, for example, a fermionic system with on-site potential and nearest-neighbor hopping terms. 
The XY model, a hard-core boson system, and the transverse-field Ising model can also be mapped to this form using the Jordan-Wigner transformation.

We first consider the case of $B_{l}=0$, for which the total particle number is conserved.
This system can be diagonalized by the Fourier transform as
\beq
H=\sum_p\varepsilon_pf_p^{\dagger}f_p,
\label{eq:H_diag}
\eeq
where $f_p^{\dagger}=(1/\sqrt{L})\sum_{x=1}^Lc_x^{\dagger}e^{-ipx}$ and $\varepsilon_p=\sum_{x=1}^LA_xe^{ipx}$.
The summation over $p=2\pi m/L$ is taken over integers $m$ with $-(L-1)/2\leq m\leq(L-1)/2$, where we consider the case of odd $L$ throughout the paper, although this restriction is not essential.

The occupation-number operator of each of the $L$ eigenmodes $\{f^{\dagger}_pf_p\}$ is a conserved quantity.
Although $f^{\dagger}_pf_p$ are not spatially local, we can construct macroscopic local conserved quantities out of them as 
\beqa
\left\{
\begin{split}
&\mathcal{Q}^{(+)}_n=\sum_p\cos{\left(np\right)}f_p^{\dag}f_p,& &n=0,1,\dots,\frac{L-1}{2},
\label{eq-Q+}
\\
&\mathcal{Q}^{(-)}_n=\sum_p\sin{\left(np\right)}f_p^{\dag}f_p,& &n=1,\dots,\frac{L-1}{2};
\label{eq-Q-}
\end{split}
\right.
\label{eq:cons}
\eeqa
see Ref.~\cite{Essler2016}.
We then define $\mathcal{Q}^{(+)}_{-n}=\mathcal{Q}^{(+)}_n$, $\mathcal{Q}^{(-)}_{-n}=-\mathcal{Q}^{(-)}_n$, and $\mathcal{Q}^{(-)}_0=0$.
Note that $\mathcal{Q}^{(+)}_0$ coincides with the total particle number: 
\beq
\mathcal{Q}^{(+)}_0=\sum_pf_p^{\dag}f_p=\hat{N}.
\eeq
We denote an eigenvalue of $\mathcal{Q}_n^{(\pm)}$ for the Fock eigenstates by $Q_n^{(\pm)}$. 

When $B_{l}\neq 0$, the Bogoliubov transformation following the Fourier transformation diagonalizes the Hamiltonian as
\beq
H=\sum_p\tilde{\varepsilon}_p\eta_p^{\dagger}\eta_p+{\rm const}.,
\label{eq:H_diag2}
\eeq
where $\tilde{\varepsilon}_p$ and $\eta_p^{\dagger}$ are given by $a_p:=\sum_{x=1}^LA_xe^{ipx}$ and $b_p:=2i\sum_{x=1}^LB_x\sin(px)$ as
\beqa
&\displaystyle{\tilde{\varepsilon}_p=\frac{a_p-a_{-p}+\sqrt{(a_p+a_{-p})^2+4|b_p|^2}}{2}},\\
&\eta_p^{\dagger}=s(p)f_p^{\dagger}+t(p)f_{-p},
\eeqa
with the functions $s(p)$ and $t(p)$ defined as
\beqa
&&s(p)=\frac{|b_p|}{\sqrt{|b_p|^2+(\tilde{\varepsilon}_p-a_p)^2}}, \\
&&t(p)=\frac{|b_p|}{b_p}\frac{\tilde{\varepsilon}_p-a_p}{\sqrt{|b_p|^2+(\tilde{\varepsilon}_p-a_p)^2}}.
\eeqa
Macroscopic local conserved quantities in this case are given by
\beq
\left\{
\begin{split}
&\mathcal{Q}_n^{(+)}=\frac{1}{2}\sum_p\cos(np)(\tilde{\varepsilon}_p+\tilde{\varepsilon}_{-p})\eta_p^{\dagger}\eta_p,
\\
&\mathcal{Q}_n^{(-)}=\sum_p\sin(np)\eta_p^{\dagger}\eta_p,
\end{split}
\right.
\label{eq:cons2}
\eeq
where we use the same notations as in Eq.~(\ref{eq:cons}), but there will be no confusion. 

The locality of $\mathcal{Q}_n^{(+)}$ in Eq.~(\ref{eq:cons2}) is proved as follows.
First, we divide it into two parts as follows:  
\beq
\mathcal{Q}_n^{(+)}=\sum_p\tilde{\varepsilon}_p\cos(np)\eta_p^\dagger\eta_p+\sum_p\frac{\tilde{\varepsilon}_p-\tilde{\varepsilon}_{-p}}{2}\cos(np)\eta_p^\dagger\eta_p.
\label{eq:cons_local}
\eeq
It is known and explicitly confirmed that the first term of Eq.~(\ref{eq:cons_local}) is local~\cite{Fagotti-Essler2013}.
As for the second term, we notice that $\tilde{\varepsilon}_p-\tilde{\varepsilon}_{-p}=a_p-a_{-p}$ is written as a finite sum $\sum_{x=-r_H}^{r_H}A_x(e^{ipx}-e^{-ipx})$ because of the fact that $\hat{H}$ is a local operator with the maximum range $r_H$.
Therefore, the second term of Eq.~(\ref{eq:cons_local}) is written as a linear combination of $\{\mathcal{Q}_m^{(-)}\}$ with $m\leq n+r_H$, which is a local operator.
Thus, for any fixed $n$, both the first and the second terms of Eq.~(\ref{eq:cons2}) are local in the thermodynamic limit.

In terms of these local conserved quantities, the GGE is given as the density matrix
\beq
\rho_\mathrm{GGE}=\frac{e^{-\sum_{n=0}^{(L-1)/2}\left(\Lambda_n^{(+)}\mathcal{Q}_n^{(+)}+\Lambda_n^{(-)}\mathcal{Q}_n^{(-)}\right)}}{Z_\mathrm{GGE}},
\eeq
where $Z_\mathrm{GGE}$ is the normalization constant.
The parameters $\Lambda_p^{(\pm)}$ are determined from the initial state $|\psi(0)\>$ by the condition that $\<\psi(0)|\mathcal{Q}_n^{(\pm)}|\psi(0)\>=\Tr[\mathcal{Q}_n^{(\pm)}\rho_\mathrm{GGE}]$.

\sectionprl{Strong generalized ETH}
\label{sec:gETH}
In order to formulate our generalized ETH, we first define a Hilbert subspace called an $n_c$-shell with the notation $\mathcal{S}_{n_c}$.
Let us denote the set of the simultaneous eigenstates of $\{\mathcal{Q}_n^{(\pm)}\}$ by $\mathcal{E}$.
An $n_c$-shell is then defined as the Hilbert subspace spanned by all the eigenstates in $\mathcal{E}$ with the eigenvalues located around the center $\{\bar{Q}_n^{(\pm)}\}_{n=1}^{n_c}$: 
\begin{align}
\mathcal{S}_{n_c}:=&\mathrm{Span}\Big\{|\alpha\>\in\mathcal{E}: \text{ for all }0\leq n\leq n_c, \nonumber\\
&\left. Q_n^{(\pm)}\in[\bar{Q}_n^{(\pm)}-\Delta_n^{(\pm)},\bar{Q}_n^{(\pm)}+\Delta_n^{(\pm)}]\right\}.
\end{align}
Here, the half width of the shell $\Delta_n^{(\pm)}$ is arbitrary as long as it is microscopically large but macroscopically small; for example, we can choose $\Delta_n\propto L^{1/2}$. 
Note that $n$ runs up to $n_c\leq(L-1)/2.$ 
The $n_c$-shell can be regarded as a generalization of the usual energy shell in the microcanonical ensemble.

Now we formulate the strong generalized ETH.
It states that all the energy eigenstates in $\mathcal{S}_{n_c}$ are locally indistinguishable from each other in the limit of $n_c\rightarrow\infty$ taken after the thermodynamic limit $L\rightarrow\infty$.
For convenience, we also say that a local observable $\hat{o}$ satisfies the $n_c$-ETH when $\<\alpha|\hat{o}|\alpha\>=\<\alpha'|\hat{o}|\alpha'\>$ for any pair of eigenstates $|\alpha\>,|\alpha'\>\in\mathcal{S}_{n_c}$ in the thermodynamic limit.

It should be noted that another generalization of ETH has been proposed in the previous work~\cite{Cassidy2011}, stating that energy eigenstates with similar distributions of the mode occupation number look similar with respect to local observables.
Below we explain the relation between our generalized ETH based on the $n_c$-shell and the generalized ETH based on the mode occupation number distibutions, which is a simplified version of the one originally proposed in Ref.~\cite{Cassidy2011}.

For simplicity, we consider the case in which the total particle number is conserved with $B_{l}=0$.
Then, each energy eigenstate $|\alpha\>$ consists of $N$ occupied levels $\{p_1^{\alpha},p_2^{\alpha},\dots,p_N^{\alpha}\}$, where $p_i^{\alpha}=2\pi n_i^{\alpha}/L$ with integers $\{n_i^{\alpha}\}_{i=1}^N$ satisfying  $-\pi\leq p_1^{\alpha}<p_2^{\alpha}<\dots<p_N^{\alpha}<\pi$. 
In short, $\<\alpha|f_p^{\dagger}f_p|\alpha\>=1$ if and only if $p\in\{p_1^{\alpha},p_2^{\alpha},\dots,p_N^{\alpha}\}$.
Let us say that two eigenstates $|\alpha\>$ and $|\alpha'\>$ have `similar' distributions of the mode occupation number if and only if
\beq
\delta(\alpha,\alpha')=\left[\frac{1}{N}\sum_{i=1}^N\left(p_i^{\alpha}-p_i^{\alpha'}\right)^2\right]^{1/2}
\eeq
is smaller than a threshold $\epsilon$, which can be set to zero in the thermodynamic limit.
The generalized ETH formulated in Ref.~\cite{Cassidy2011} essentially states that two eigenstates with similar distributions of the mode occupation number are locally indistinguishable. 
Now we begin the explanation of its relation with our generalized ETH. 
Let us consider the difference of a macroscopic conserved quantity in the states $|\alpha\>$ and $|\alpha'\>$:
\beq
\delta q_n^{(\pm)}:=\frac{1}{L}\left|\<\alpha|\mathcal{Q}_n^{(\pm)}|\alpha\>-\<\alpha'|\mathcal{Q}_n^{(\pm)}|\alpha'\>\right|.
\eeq
If $|\delta q_n^{(\pm)}|\leq2\Delta_n^{(\pm)}/L$ for all $n\leq n_c$, the two eigenstates $|\alpha\>$ and $|\alpha'\>$ belong to the same $n_c$-shell under a suitable choice of the center of the shell $\{\bar{Q}_n^{(\pm)}\}_{n=1}^{n_c}$.
By using $p_i^{\alpha}$, we can rewrite $\delta q_n^{(+)}$ as
\begin{align}
\delta q_n^{(+)}=\frac{1}{L}\left|\sum_{i=1}^N[\cos(np_i^{\alpha})-\cos(np_i^{\alpha'})]\right|
\nonumber \\
\leq\frac{1}{L}\sum_{i=1}^N\left|\cos(np_i^{\alpha})-\cos(np_i^{\alpha'})\right|
\end{align}
By using $|\cos\theta-\cos\phi|\leq|\theta-\phi|$, we obtain
\begin{align}
\delta q_n^{(+)}&\leq\frac{n}{L}\sum_{i=1}^N|p_i^{\alpha}-p_i^{\alpha'}|
\nonumber \\
&\leq n\rho\delta(\alpha,\alpha'),
\end{align}
where $\rho=N/L$ and we have used $\delta(\alpha,\alpha')\geq(1/N)\sum_{i=1}^N|p_i^{\alpha}-p_i^{\alpha'}|$.
Similarly, $\delta q_n^{(-)}\leq n\rho\delta(\alpha,\alpha')$ holds. 

From these inequalities, we can immediately conclude that two eigenstates $|\alpha\>$ and $|\alpha'\>$ belong to the same $n_c$-shell under a suitable choice of $\{\bar{Q}_n^{(\pm)}\}_{n=1}^{n_c}$ as long as $\delta(\alpha,\alpha')\leq2\Delta_n^{(\pm)}/(n_cN)$.
Since $\Delta_n^{(\pm)}$ is chosen so that $\Delta_n^{(\pm)}/N\rightarrow 0$ in the thermodynamic limit, this result implies that two eigenstates $|\alpha\>$ and $|\alpha'\>$ with similar distributions of the mode occupation number belong to the same $n_c$-shell.
This implies that if the generalized ETH based on the $n_c$-shell holds in the strong sense, then the generalized ETH based on the similarity of the distributions of the mode occupation number also holds in the strong sense~\footnote
{It should be noted that the converse is not true in general.}.
Thus the proof of the strong generalized ETH based on the $n_c$-shell complements the numerical result in Ref.~\cite{Cassidy2011}, in which the generalized ETH based on the mode occupation number distribution has been confirmed only in the weak sense.

\sectionprl{Proof}
\label{sec:proof}
We consider local observables $\hat{o}$ which consists of fermionic operators $\{c^{\dag},c\}$ with the maximum range $r$. 
For example, 
\beq
\hat{o}=\frac{1}{L}\sum_{j=1}^L\left(c^{\dag}_{j+2}c_{j+2}c^{\dag}_jc_j+c^{\dag}_{j+1}c_j+c^{\dag}_jc_{j+1}\right)
\eeq
is the case of $r=2$.
As a shorthand notation, we write $\<\hat{o}\>:=\<\alpha|\hat{o}|\alpha\>$ for a fixed eigenstate $|\alpha\>$.


We first consider the case $B_{l}=0$, in which the total particle number is conserved. 
In this case, the diagonalized Hamiltonian is given by Eq.~(\ref{eq:H_diag}) and macroscopic conserved quantities are given by Eq.~(\ref{eq:cons}).
We shall prove that the eigenstate expectation value of a local observable can be written as a smooth function of the eigenvalues $\{Q^{(\pm)}_m/L\}$ of the constructed conserved quantities with $m\leq r$.
In other words, any local observable with the maximum range $r$ satisfies the $r$-ETH.

By virtue of Wick's theorem (see the note in~\footnote{Although Wick's theorem is usually applied to the vacuum state, it can be applied to individual eigenstates too in the present systems because the eigenstates can be expressed as a vacuum state by redefining the particles and holes for each mode. 
See Ref.~\cite{Molinari_arXiv2017}.}), 
the eigenstate expectation value $\<\hat{o}\>$ of any local observable $\hat{o}$ with the maximum range $r$ can be decomposed into products of two-point functions of the form $\<c_x^{\dag}c_y\>$ with $|x-y|_P\leq r$.
More precisely, if we denote by $X_i$ a linear superposition of $\{c_x,c_x^\dagger\}$,
\begin{align}
&\<X_1X_2\dots X_{2n}\>\nonumber \\
&=\sum(-1)^P\<X_{i_1}X_{j_1}\>\<X_{i_2}X_{j_2}\>\dots\<X_{i_n}X_{j_n}\>,
\label{eq:Wick}
\end{align}
where the sum is over all partitions of $1,2,\dots 2n$ into pairs $\{(i_1,j_1),(i_2,j_2),\dots,(i_n,j_n)\}$ with $i_1<j_1$, $i_2<j_2,\dots i_n<j_n$, and $P$ is the parity of the permutation $(1,2,\dots,2n)\rightarrow (i_1,j_1,i_2,j_2,\dots,i_n,j_n)$~\cite{Molinari_arXiv2017}.
It should be noted that Eq.~(\ref{eq:Wick}) also holds even when $B_{l}\neq 0$ because the Bogoliubov fermion operators $\eta_p$ and $\eta_p^\dagger$ can be written as a linear superposition of $\{c_x,c_x^\dagger\}$.

We can express the two-point function in terms of the conserved quantities in Eq.~\eqref{eq:cons} as in 
\beqa
\<c_x^{\dag}c_y\>
=\frac{1}{L}\sum_{p}e^{ip(x-y)}\<f_p^{\dag}f_p\>\nonumber\\
=\frac{1}{L}(Q_{x-y}^{(+)}+iQ_{x-y}^{(-)}).
\eeqa
Therefore, $\<\hat{o}\>$ is generally a smooth function of $\{Q^{(\pm)}_m/L\}$ with $m\leq r$.  

This immediately leads to the validity of the strong generalized ETH.
Moreover, any local operator with the maximum range $r\leq n_c$ satisfies the $n_c$-ETH.
Therefore, as far as we consider local operators with a fixed maximum range $r$, the steady state is described by the microcanonical ensemble within the $n_c$-shell, which is in the thermodynamic limit equivalent to the \textit{truncated} GGE,
\beq
\rho_\mathrm{GGE}^{(n_c)}:=\frac{\exp\left[-\sum_{n=0}^{n_c}\left(\Lambda_n^{(+)}\mathcal{Q}_n^{(+)}+\Lambda_n^{(-)}\mathcal{Q}_n^{(-)}\right)\right]}{Z_\mathrm{GGE}^{(n_c)}},
\eeq
for an arbitrary $n_c\geq r$, where $Z_\mathrm{GGE}^{(n_c)}$ is the normalization factor.
In the limit of $n_c\rightarrow\infty$ after the thermodynamic limit, the GGE reproduces expectation values of arbitrary local operators in the steady state.


Next, we consider free fermion models in which the total particle number is not conserved ($B_{l}\neq 0$).
The eigenstate expectation value of a local operator is again decomposed into the products of two-point functions.
Relevant two-point functions are $\<c_x^{\dagger}c_y\>$ and $\<c_x^{\dagger}c_y^{\dagger}\>$ with $|x-y|_P\leq r,$ the latter of which appears because $B_{x-y}\neq 0$.
By expressing these two-point functions using the mode occupation numbers $\eta_p^{\dagger}\eta_p$, we have
\begin{align}
\<c_x^{\dagger}c_y\>=&\frac{1}{L}\sum_p\cos[p(x-y)]\left(s(p)^2-|t(p)|^2\right)\<\eta_p^{\dagger}\eta_p\>
\nonumber \\
&+\frac{i}{L}\sum_p\sin[p(x-y)]\<\eta_p^{\dagger}\eta_p\>+\mathrm{const.},
\end{align}
and
\begin{align}
\<c_x^{\dagger}c_y^{\dagger}\>=\frac{2i}{L}\sum_p\sin[p(x-y)]s(p)t(p)\<\eta_p^{\dagger}\eta_p\>+\mathrm{const.}
\end{align}
By performing the Fourier series expansion, we can express $\<c_x^{\dagger}c_y\>$ and $\<c_x^{\dagger}c_y^{\dagger}\>$ as
\begin{align}
\<c_x^{\dagger}c_y\>=\frac{v_0}{L}Q_{x-y}^{(+)}&+\frac{1}{L}\sum_{n=1}^{(L-1)/2}v_n\left(Q_{x-y+n}^{(+)}+Q_{x-y-n}^{(+)}\right)
\nonumber \\
&+\frac{i}{L}Q_{x-y}^{(-)}+\mathrm{const.},
\label{eq:cdxcy}
\end{align}
and
\begin{align}
\<c_x^{\dagger}c_y^{\dagger}\>=-\frac{i}{L}\sum_{n=1}^{(L-1)/2}w_n\left(Q_{x-y+n}^{(+)}-Q_{x-y-n}^{(+)}\right)+\mathrm{const.}
\label{eq:cdxcdy}
\end{align}
Here, $v_n$ in Eq.~(\ref{eq:cdxcy}) and $w_n$ in Eq.~(\ref{eq:cdxcdy}) are the Fourier coefficients of
\beq
\frac{2}{\tilde{\varepsilon}(p)+\tilde{\varepsilon}(-p)}\left(s(p)^2-|t(p)|^2\right)
\label{eq:fv}
\eeq
and
\beq
\frac{4i}{\tilde{\varepsilon}(p)+\tilde{\varepsilon}(-p)}s(p)t(p),
\label{eq:fw}
\eeq
respectively, where the Fourier coefficient of $\phi(p)$ is defined by $\phi_n=(1/L)\sum_p\phi(p)e^{-ipn}$.
It is noted that the relations $v_n=v_{-n}$ and $w_n=-w_{-n}$, which follows from the parity of the functions \eqref{eq:fv} and \eqref{eq:fw}, are used in deriving Eqs.~(\ref{eq:cdxcy}) and (\ref{eq:cdxcdy}).
According to the Riemann-Lebesgue lemma, $v_n$ and $w_n$ tend to zero in the limit of $|n|\rightarrow\infty$ taken after the thermodynamic limit (see the note in~\footnote{The value $\tilde{\varepsilon}(p)+\tilde{\varepsilon}(-p)$ is non-negative for all $p$. 
In the case in which $\tilde{\varepsilon}(p)+\tilde{\varepsilon}(-p)$ touches the $p$-axis, the functions \eqref{eq:fv} and \eqref{eq:fw} may not be $L^1$-integrable. 
In this case, the following arguments remain valid by replacing $\tilde{\varepsilon}(p)+\tilde{\varepsilon}(-p)$ by a constant $\delta_c>0$ when $\tilde{\varepsilon}(p)+\tilde{\varepsilon}(-p)$ becomes smaller than $\delta_c$, and taking the limit $\delta_c\to 0$ after $L\to \infty$ and $n_c\to \infty$}).
Therefore, we can approximately truncate the summations over $n$ in Eqs.~(\ref{eq:cdxcy}) and (\ref{eq:cdxcdy}) at a sufficiently large $n^*$, e.g.,
\begin{align}
\<c_x^{\dagger}c_y\>\approx\frac{v_0}{L}Q_{x-y}^{(+)}&+\frac{1}{L}\sum_{n=1}^{n^*}v_n\left(Q_{x-y+n}^{(+)}+Q_{x-y-n}^{(+)}\right)
\nonumber \\
&+\frac{i}{L}Q_{x-y}^{(-)}+\mathrm{const.}
\label{eq:cdxcyap}
\end{align}
This approximation becomes exact in the limit of $n^*\rightarrow\infty$ taken after the thermodynamic limit.

In this way, the eigenstate expectation value of a local operator with a maximum range $r$ is approximately written as a linear combination of $Q_n^{(\pm)}/L$ with $n\leq r+n^*$, and this approximation becomes exact in the limit of $n^*\rightarrow\infty$.
It implies that any local operator satisfies $n_c$-ETH in the limit of $n_c\rightarrow\infty$ after the thermodynamic limit.
Thus, the strong generalized ETH has been proved.

\sectionprl{Conclusion}
\label{sec:conclusion}
The strong generalized ETH proved in this work ensures that if the initial state is in a generalized shell constructed by local conserved quantities, the system relaxes to a steady state that is described by the GGE, either truncated or not.
Since a physically relevant initial state, e.g., a state prepared by a quench, has subextensive fluctuations of macroscopic quantities, such an initial state is necessarily in a generalized shell.
Therefore, a steady state after relaxation is described by a GGE in a translationally invariant noninteracting integrable system.
Our results can be generalized to $d$-dimensional systems and noninteracting bosons.

In the previous studies, the validity of the GGE has been proved for noninteracting integrable models with translation invariance by requiring the cluster decomposition property for the initial state~\cite{Cramer2008,Sotiriadis2014,Gluza2016}. 
In contrast, our result applies to dynamics with the initial state which can be any state in a single generalized shell. 
Since the cluster decomposition property does not hold for all of such states, our result shows that the GGE is valid for a wider class of initial states than expected previously.
It should be noted that the removal of the assumption of the cluster decomposition property is particularly important when we consider a spin model that is mapped to quadratic fermions, e.g., the transverse-field Ising chain and the XY chain.
In these models, a physically realistic initial state should obey the cluster decomposition property \textit{with respect to the spin operators}, but it is not obvious whether the same initial state obeys the cluster decomposition property \textit{with respect to the fermion operators}~\cite{Murthy_arXiv2018}.

In this work, we have assumed the translation invariance and the locality of the quadratic Hamiltonian.
It is a future problem to clarify whether these assumptions are essential for the relaxation towards a GGE.
Considering the case of nonlocal Hamiltonians is important in validating the relaxation to the Floquet GGE~\cite{Lazarides2014a, Ishii-Kuwahara2018} in the low-frequency regime of time-periodic systems, where the effective Hamiltonian generally becomes nonlocal~\cite{DAlessio2014}. 

It is also open to extend the strong generalized ETH to interacting integrable systems, which cannot be mapped to free particles but exactly solvable via the method of the Bethe ansatz.
Although the idea of the generalized ETH has been applied to interacting integrable systems~\cite{Caux2013,Ilievski2016}, it has not been proven in the strong sense.
A recent finding of quasi-local charges in the XXZ chain has advanced our understanding on the validity of the GGE in interacting integrable systems~\cite{Ilievski2015}.
The quasi-local charges should be taken into account properly; otherwise, the generalized ETH cannot be true~\cite{Goldstein2014,Pozsgay2014} and the GGE fails~\cite{Wouters2014, Pozsgay2014_correlations,Brockmann2014, Mestyan2015}.

\begin{acknowledgments}
We are grateful to Professor N. Hatano for fruitful discussions and careful reading of the manuscript.
We also thank Dr. T. N. Ikeda and Dr. Y. Watanabe for useful discussions. 
We thank Professor M. Rigol for valuable comments on the manuscript. 
T.I. was supported by the Program for Leading Graduate Schools, MEXT, Japan. 
\end{acknowledgments}

\bibliography{vge.bib}

\begin{thebibliography}{60}%
\makeatletter
\providecommand \@ifxundefined [1]{%
 \@ifx{#1\undefined}
}%
\providecommand \@ifnum [1]{%
 \ifnum #1\expandafter \@firstoftwo
 \else \expandafter \@secondoftwo
 \fi
}%
\providecommand \@ifx [1]{%
 \ifx #1\expandafter \@firstoftwo
 \else \expandafter \@secondoftwo
 \fi
}%
\providecommand \natexlab [1]{#1}%
\providecommand \enquote  [1]{``#1''}%
\providecommand \bibnamefont  [1]{#1}%
\providecommand \bibfnamefont [1]{#1}%
\providecommand \citenamefont [1]{#1}%
\providecommand \href@noop [0]{\@secondoftwo}%
\providecommand \href [0]{\begingroup \@sanitize@url \@href}%
\providecommand \@href[1]{\@@startlink{#1}\@@href}%
\providecommand \@@href[1]{\endgroup#1\@@endlink}%
\providecommand \@sanitize@url [0]{\catcode `\\12\catcode `\$12\catcode
  `\&12\catcode `\#12\catcode `\^12\catcode `\_12\catcode `\%12\relax}%
\providecommand \@@startlink[1]{}%
\providecommand \@@endlink[0]{}%
\providecommand \url  [0]{\begingroup\@sanitize@url \@url }%
\providecommand \@url [1]{\endgroup\@href {#1}{\urlprefix }}%
\providecommand \urlprefix  [0]{URL }%
\providecommand \Eprint [0]{\href }%
\providecommand \doibase [0]{http://dx.doi.org/}%
\providecommand \selectlanguage [0]{\@gobble}%
\providecommand \bibinfo  [0]{\@secondoftwo}%
\providecommand \bibfield  [0]{\@secondoftwo}%
\providecommand \translation [1]{[#1]}%
\providecommand \BibitemOpen [0]{}%
\providecommand \bibitemStop [0]{}%
\providecommand \bibitemNoStop [0]{.\EOS\space}%
\providecommand \EOS [0]{\spacefactor3000\relax}%
\providecommand \BibitemShut  [1]{\csname bibitem#1\endcsname}%
\let\auto@bib@innerbib\@empty
\bibitem [{\citenamefont {D'Alessio}\ \emph {et~al.}(2016)\citenamefont
  {D'Alessio}, \citenamefont {Kafri}, \citenamefont {Polkovnikov},\ and\
  \citenamefont {Rigol}}]{DAlessio_2016}%
  \BibitemOpen
  \bibfield  {author} {\bibinfo {author} {\bibfnamefont {L.}~\bibnamefont
  {D'Alessio}}, \bibinfo {author} {\bibfnamefont {Y.}~\bibnamefont {Kafri}},
  \bibinfo {author} {\bibfnamefont {A.}~\bibnamefont {Polkovnikov}}, \ and\
  \bibinfo {author} {\bibfnamefont {M.}~\bibnamefont {Rigol}},\ }\href
  {https://doi.org/10.1080/00018732.2016.1198134} {\bibfield  {journal}
  {\bibinfo  {journal} {Advances in Physics}\ }\textbf {\bibinfo {volume}
  {65}},\ \bibinfo {pages} {239} (\bibinfo {year} {2016})}\BibitemShut
  {NoStop}%
\bibitem [{\citenamefont {Eisert}\ \emph {et~al.}(2015)\citenamefont {Eisert},
  \citenamefont {Friesdorf},\ and\ \citenamefont
  {Gogolin}}]{Eisert_review2015}%
  \BibitemOpen
  \bibfield  {author} {\bibinfo {author} {\bibfnamefont {J.}~\bibnamefont
  {Eisert}}, \bibinfo {author} {\bibfnamefont {M.}~\bibnamefont {Friesdorf}}, \
  and\ \bibinfo {author} {\bibfnamefont {C.}~\bibnamefont {Gogolin}},\ }\href
  {\doibase 10.1038/nphys3215} {\bibfield  {journal} {\bibinfo  {journal} {Nat.
  Phys.}\ }\textbf {\bibinfo {volume} {11}},\ \bibinfo {pages} {124} (\bibinfo
  {year} {2015})}\BibitemShut {NoStop}%
\bibitem [{\citenamefont {Mori}\ \emph {et~al.}(2018)\citenamefont {Mori},
  \citenamefont {Ikeda}, \citenamefont {Kaminishi},\ and\ \citenamefont
  {Ueda}}]{mori2018review}%
  \BibitemOpen
  \bibfield  {author} {\bibinfo {author} {\bibfnamefont {T.}~\bibnamefont
  {Mori}}, \bibinfo {author} {\bibfnamefont {T.~N.}\ \bibnamefont {Ikeda}},
  \bibinfo {author} {\bibfnamefont {E.}~\bibnamefont {Kaminishi}}, \ and\
  \bibinfo {author} {\bibfnamefont {M.}~\bibnamefont {Ueda}},\ }\href {\doibase
  10.1088/1361-6455/aabcdf} {\bibfield  {journal} {\bibinfo  {journal} {J.
  Phys. B}\ }\textbf {\bibinfo {volume} {51}},\ \bibinfo {pages} {112001}
  (\bibinfo {year} {2018})}\BibitemShut {NoStop}%
\bibitem [{\citenamefont {Trotzky}\ \emph {et~al.}(2012)\citenamefont
  {Trotzky}, \citenamefont {Chen}, \citenamefont {Flesch}, \citenamefont
  {Mcculloch}, \citenamefont {Schollw{\"{o}}ck}, \citenamefont {Eisert},\ and\
  \citenamefont {Bloch}}]{Trotzky2012}%
  \BibitemOpen
  \bibfield  {author} {\bibinfo {author} {\bibfnamefont {S.}~\bibnamefont
  {Trotzky}}, \bibinfo {author} {\bibfnamefont {Y.-A.}\ \bibnamefont {Chen}},
  \bibinfo {author} {\bibfnamefont {A.}~\bibnamefont {Flesch}}, \bibinfo
  {author} {\bibfnamefont {I.~P.}\ \bibnamefont {Mcculloch}}, \bibinfo {author}
  {\bibfnamefont {U.}~\bibnamefont {Schollw{\"{o}}ck}}, \bibinfo {author}
  {\bibfnamefont {J.}~\bibnamefont {Eisert}}, \ and\ \bibinfo {author}
  {\bibfnamefont {I.}~\bibnamefont {Bloch}},\ }\href {\doibase
  10.1038/NPHYS2232} {\bibfield  {journal} {\bibinfo  {journal} {Nat. Phys.}\
  }\textbf {\bibinfo {volume} {8}},\ \bibinfo {pages} {325} (\bibinfo {year}
  {2012})}\BibitemShut {NoStop}%
\bibitem [{\citenamefont {Kaufman}\ \emph {et~al.}(2016)\citenamefont
  {Kaufman}, \citenamefont {Tai}, \citenamefont {Lukin}, \citenamefont
  {Rispoli}, \citenamefont {Schittko}, \citenamefont {Preiss},\ and\
  \citenamefont {Greiner}}]{Kaufman2016}%
  \BibitemOpen
  \bibfield  {author} {\bibinfo {author} {\bibfnamefont {A.~M.}\ \bibnamefont
  {Kaufman}}, \bibinfo {author} {\bibfnamefont {M.~E.}\ \bibnamefont {Tai}},
  \bibinfo {author} {\bibfnamefont {A.}~\bibnamefont {Lukin}}, \bibinfo
  {author} {\bibfnamefont {M.}~\bibnamefont {Rispoli}}, \bibinfo {author}
  {\bibfnamefont {R.}~\bibnamefont {Schittko}}, \bibinfo {author}
  {\bibfnamefont {P.~M.}\ \bibnamefont {Preiss}}, \ and\ \bibinfo {author}
  {\bibfnamefont {M.}~\bibnamefont {Greiner}},\ }\href
  {http://science.sciencemag.org/content/353/6301/794/tab-pdf} {\bibfield
  {journal} {\bibinfo  {journal} {Science}\ }\textbf {\bibinfo {volume}
  {353}},\ \bibinfo {pages} {794} (\bibinfo {year} {2016})}\BibitemShut
  {NoStop}%
\bibitem [{\citenamefont {Neill}\ \emph {et~al.}(2016)\citenamefont {Neill},
  \citenamefont {Roushan}, \citenamefont {Fang}, \citenamefont {Chen},
  \citenamefont {Kolodrubetz}, \citenamefont {Chen}, \citenamefont {Megrant},
  \citenamefont {Barends}, \citenamefont {Campbell}, \citenamefont {Chiaro},
  \citenamefont {Dunsworth}, \citenamefont {Jeerey}, \citenamefont {Kelly},
  \citenamefont {Mutus}, \citenamefont {{O 'malley}}, \citenamefont {Quintana},
  \citenamefont {Sank}, \citenamefont {Vainsencher}, \citenamefont {Wenner},
  \citenamefont {White}, \citenamefont {Polkovnikov},\ and\ \citenamefont
  {Martinis}}]{Neill2016}%
  \BibitemOpen
  \bibfield  {author} {\bibinfo {author} {\bibfnamefont {C.}~\bibnamefont
  {Neill}}, \bibinfo {author} {\bibfnamefont {P.}~\bibnamefont {Roushan}},
  \bibinfo {author} {\bibfnamefont {M.}~\bibnamefont {Fang}}, \bibinfo {author}
  {\bibfnamefont {Y.}~\bibnamefont {Chen}}, \bibinfo {author} {\bibfnamefont
  {M.}~\bibnamefont {Kolodrubetz}}, \bibinfo {author} {\bibfnamefont
  {Z.}~\bibnamefont {Chen}}, \bibinfo {author} {\bibfnamefont {A.}~\bibnamefont
  {Megrant}}, \bibinfo {author} {\bibfnamefont {R.}~\bibnamefont {Barends}},
  \bibinfo {author} {\bibfnamefont {B.}~\bibnamefont {Campbell}}, \bibinfo
  {author} {\bibfnamefont {B.}~\bibnamefont {Chiaro}}, \bibinfo {author}
  {\bibfnamefont {A.}~\bibnamefont {Dunsworth}}, \bibinfo {author}
  {\bibfnamefont {E.}~\bibnamefont {Jeerey}}, \bibinfo {author} {\bibfnamefont
  {J.}~\bibnamefont {Kelly}}, \bibinfo {author} {\bibfnamefont
  {J.}~\bibnamefont {Mutus}}, \bibinfo {author} {\bibfnamefont {P.~J.~J.}\
  \bibnamefont {{O 'malley}}}, \bibinfo {author} {\bibfnamefont
  {C.}~\bibnamefont {Quintana}}, \bibinfo {author} {\bibfnamefont
  {D.}~\bibnamefont {Sank}}, \bibinfo {author} {\bibfnamefont {A.}~\bibnamefont
  {Vainsencher}}, \bibinfo {author} {\bibfnamefont {J.}~\bibnamefont {Wenner}},
  \bibinfo {author} {\bibfnamefont {T.~C.}\ \bibnamefont {White}}, \bibinfo
  {author} {\bibfnamefont {A.}~\bibnamefont {Polkovnikov}}, \ and\ \bibinfo
  {author} {\bibfnamefont {J.~M.}\ \bibnamefont {Martinis}},\ }\href {\doibase
  10.1038/NPHYS3830} {\bibfield  {journal} {\bibinfo  {journal} {Nat. Phys.}\
  }\textbf {\bibinfo {volume} {12}},\ \bibinfo {pages} {1037} (\bibinfo {year}
  {2016})}\BibitemShut {NoStop}%
\bibitem [{\citenamefont {Clos}\ \emph {et~al.}(2016)\citenamefont {Clos},
  \citenamefont {Porras}, \citenamefont {Warring},\ and\ \citenamefont
  {Schaetz}}]{Clos2016}%
  \BibitemOpen
  \bibfield  {author} {\bibinfo {author} {\bibfnamefont {G.}~\bibnamefont
  {Clos}}, \bibinfo {author} {\bibfnamefont {D.}~\bibnamefont {Porras}},
  \bibinfo {author} {\bibfnamefont {U.}~\bibnamefont {Warring}}, \ and\
  \bibinfo {author} {\bibfnamefont {T.}~\bibnamefont {Schaetz}},\ }\href
  {\doibase 10.1103/PhysRevLett.117.170401} {\bibfield  {journal} {\bibinfo
  {journal} {Phys. Rev. Lett.}\ }\textbf {\bibinfo {volume} {117}},\ \bibinfo
  {pages} {170401} (\bibinfo {year} {2016})}\BibitemShut {NoStop}%
\bibitem [{\citenamefont {Saito}\ \emph {et~al.}(1996)\citenamefont {Saito},
  \citenamefont {Takesue},\ and\ \citenamefont {Miyashita}}]{Saito1996}%
  \BibitemOpen
  \bibfield  {author} {\bibinfo {author} {\bibfnamefont {K.}~\bibnamefont
  {Saito}}, \bibinfo {author} {\bibfnamefont {S.}~\bibnamefont {Takesue}}, \
  and\ \bibinfo {author} {\bibfnamefont {S.}~\bibnamefont {Miyashita}},\ }\href
  {\doibase 10.1143/JPSJ.65.1243} {\bibfield  {journal} {\bibinfo  {journal}
  {J. Phys. Soc. Jpn.}\ }\textbf {\bibinfo {volume} {65}},\ \bibinfo {pages}
  {1243} (\bibinfo {year} {1996})}\BibitemShut {NoStop}%
\bibitem [{\citenamefont {Rigol}\ \emph {et~al.}(2008)\citenamefont {Rigol},
  \citenamefont {Dunjko},\ and\ \citenamefont {Olshanii}}]{Rigol2008}%
  \BibitemOpen
  \bibfield  {author} {\bibinfo {author} {\bibfnamefont {M.}~\bibnamefont
  {Rigol}}, \bibinfo {author} {\bibfnamefont {V.}~\bibnamefont {Dunjko}}, \
  and\ \bibinfo {author} {\bibfnamefont {M.}~\bibnamefont {Olshanii}},\ }\href
  {\doibase 10.1038/nature06838} {\bibfield  {journal} {\bibinfo  {journal}
  {Nature}\ }\textbf {\bibinfo {volume} {452}},\ \bibinfo {pages} {854}
  (\bibinfo {year} {2008})}\BibitemShut {NoStop}%
\bibitem [{\citenamefont {Jin}\ \emph {et~al.}(2010)\citenamefont {Jin},
  \citenamefont {De~Raedt}, \citenamefont {Yuan}, \citenamefont
  {I.~Katsnelson}, \citenamefont {Miyashita},\ and\ \citenamefont
  {Michielsen}}]{Jin2010}%
  \BibitemOpen
  \bibfield  {author} {\bibinfo {author} {\bibfnamefont {F.}~\bibnamefont
  {Jin}}, \bibinfo {author} {\bibfnamefont {H.}~\bibnamefont {De~Raedt}},
  \bibinfo {author} {\bibfnamefont {S.}~\bibnamefont {Yuan}}, \bibinfo {author}
  {\bibfnamefont {M.}~\bibnamefont {I.~Katsnelson}}, \bibinfo {author}
  {\bibfnamefont {S.}~\bibnamefont {Miyashita}}, \ and\ \bibinfo {author}
  {\bibfnamefont {K.}~\bibnamefont {Michielsen}},\ }\href {\doibase
  10.1143/JPSJ.79.124005} {\bibfield  {journal} {\bibinfo  {journal} {J. Phys.
  Soc. Jpn.}\ }\textbf {\bibinfo {volume} {79}},\ \bibinfo {pages} {124005}
  (\bibinfo {year} {2010})}\BibitemShut {NoStop}%
\bibitem [{\citenamefont {Deutsch}(1991)}]{Deutsch1991}%
  \BibitemOpen
  \bibfield  {author} {\bibinfo {author} {\bibfnamefont {J.~M.}\ \bibnamefont
  {Deutsch}},\ }\href {\doibase 10.1103/PhysRevA.43.2046} {\bibfield  {journal}
  {\bibinfo  {journal} {Phys. Rev. A}\ }\textbf {\bibinfo {volume} {43}},\
  \bibinfo {pages} {2046} (\bibinfo {year} {1991})}\BibitemShut {NoStop}%
\bibitem [{\citenamefont {Srednicki}(1994)}]{Srednicki1994}%
  \BibitemOpen
  \bibfield  {author} {\bibinfo {author} {\bibfnamefont {M.}~\bibnamefont
  {Srednicki}},\ }\href {\doibase 10.1103/PhysRevE.50.888} {\bibfield
  {journal} {\bibinfo  {journal} {Phys. Rev. E}\ }\textbf {\bibinfo {volume}
  {50}},\ \bibinfo {pages} {888} (\bibinfo {year} {1994})}\BibitemShut
  {NoStop}%
\bibitem [{\citenamefont {Biroli}\ \emph {et~al.}(2010)\citenamefont {Biroli},
  \citenamefont {Kollath},\ and\ \citenamefont {L\"auchli}}]{Biroli2010}%
  \BibitemOpen
  \bibfield  {author} {\bibinfo {author} {\bibfnamefont {G.}~\bibnamefont
  {Biroli}}, \bibinfo {author} {\bibfnamefont {C.}~\bibnamefont {Kollath}}, \
  and\ \bibinfo {author} {\bibfnamefont {A.~M.}\ \bibnamefont {L\"auchli}},\
  }\href {\doibase 10.1103/PhysRevLett.105.250401} {\bibfield  {journal}
  {\bibinfo  {journal} {Phys. Rev. Lett.}\ }\textbf {\bibinfo {volume} {105}},\
  \bibinfo {pages} {250401} (\bibinfo {year} {2010})}\BibitemShut {NoStop}%
\bibitem [{\citenamefont {Steinigeweg}\ \emph {et~al.}(2013)\citenamefont
  {Steinigeweg}, \citenamefont {Herbrych},\ and\ \citenamefont
  {Prelov{\v{s}}ek}}]{Steinigeweg2013}%
  \BibitemOpen
  \bibfield  {author} {\bibinfo {author} {\bibfnamefont {R.}~\bibnamefont
  {Steinigeweg}}, \bibinfo {author} {\bibfnamefont {J.}~\bibnamefont
  {Herbrych}}, \ and\ \bibinfo {author} {\bibfnamefont {P.}~\bibnamefont
  {Prelov{\v{s}}ek}},\ }\href {\doibase 10.1103/PhysRevE.87.012118} {\bibfield
  {journal} {\bibinfo  {journal} {Phys. Rev. E}\ }\textbf {\bibinfo {volume}
  {87}},\ \bibinfo {pages} {012118} (\bibinfo {year} {2013})}\BibitemShut
  {NoStop}%
\bibitem [{\citenamefont {Kim}\ \emph {et~al.}(2014)\citenamefont {Kim},
  \citenamefont {Ikeda},\ and\ \citenamefont {Huse}}]{Kim2014}%
  \BibitemOpen
  \bibfield  {author} {\bibinfo {author} {\bibfnamefont {H.}~\bibnamefont
  {Kim}}, \bibinfo {author} {\bibfnamefont {T.~N.}\ \bibnamefont {Ikeda}}, \
  and\ \bibinfo {author} {\bibfnamefont {D.~A.}\ \bibnamefont {Huse}},\ }\href
  {\doibase 10.1103/PhysRevE.90.052105} {\bibfield  {journal} {\bibinfo
  {journal} {Phys. Rev. E}\ }\textbf {\bibinfo {volume} {90}},\ \bibinfo
  {pages} {052105} (\bibinfo {year} {2014})}\BibitemShut {NoStop}%
\bibitem [{\citenamefont {Beugeling}\ \emph {et~al.}(2014)\citenamefont
  {Beugeling}, \citenamefont {Moessner},\ and\ \citenamefont
  {Haque}}]{Beugeling2014}%
  \BibitemOpen
  \bibfield  {author} {\bibinfo {author} {\bibfnamefont {W.}~\bibnamefont
  {Beugeling}}, \bibinfo {author} {\bibfnamefont {R.}~\bibnamefont {Moessner}},
  \ and\ \bibinfo {author} {\bibfnamefont {M.}~\bibnamefont {Haque}},\ }\href
  {\doibase 10.1103/PhysRevE.89.042112} {\bibfield  {journal} {\bibinfo
  {journal} {Phys. Rev. E}\ }\textbf {\bibinfo {volume} {89}},\ \bibinfo
  {pages} {042112} (\bibinfo {year} {2014})}\BibitemShut {NoStop}%
\bibitem [{\citenamefont {Shiraishi}\ and\ \citenamefont
  {Mori}(2017)}]{Shiraishi-Mori2017}%
  \BibitemOpen
  \bibfield  {author} {\bibinfo {author} {\bibfnamefont {N.}~\bibnamefont
  {Shiraishi}}\ and\ \bibinfo {author} {\bibfnamefont {T.}~\bibnamefont
  {Mori}},\ }\href {\doibase 10.1103/PhysRevLett.119.030601} {\bibfield
  {journal} {\bibinfo  {journal} {Phys. Rev. Lett.}\ }\textbf {\bibinfo
  {volume} {119}},\ \bibinfo {pages} {030601} (\bibinfo {year}
  {2017})}\BibitemShut {NoStop}%
\bibitem [{\citenamefont {Mori}\ and\ \citenamefont
  {Shiraishi}(2017)}]{Mori-Shiraishi2017}%
  \BibitemOpen
  \bibfield  {author} {\bibinfo {author} {\bibfnamefont {T.}~\bibnamefont
  {Mori}}\ and\ \bibinfo {author} {\bibfnamefont {N.}~\bibnamefont
  {Shiraishi}},\ }\href {\doibase 10.1103/PhysRevE.96.022153} {\bibfield
  {journal} {\bibinfo  {journal} {Phys. Rev. E}\ }\textbf {\bibinfo {volume}
  {96}},\ \bibinfo {pages} {022153} (\bibinfo {year} {2017})}\BibitemShut
  {NoStop}%
\bibitem [{\citenamefont {Mori}(2016)}]{Mori2016}%
  \BibitemOpen
  \bibfield  {author} {\bibinfo {author} {\bibfnamefont {T.}~\bibnamefont
  {Mori}},\ }\href {https://arxiv.org/abs/1609.09776} {\bibfield  {journal}
  {\bibinfo  {journal} {arXiv:1609.09776}\ } (\bibinfo {year}
  {2016})}\BibitemShut {NoStop}%
\bibitem [{\citenamefont {Iyoda}\ \emph {et~al.}(2017)\citenamefont {Iyoda},
  \citenamefont {Kaneko},\ and\ \citenamefont {Sagawa}}]{Iyoda2017}%
  \BibitemOpen
  \bibfield  {author} {\bibinfo {author} {\bibfnamefont {E.}~\bibnamefont
  {Iyoda}}, \bibinfo {author} {\bibfnamefont {K.}~\bibnamefont {Kaneko}}, \
  and\ \bibinfo {author} {\bibfnamefont {T.}~\bibnamefont {Sagawa}},\ }\href
  {\doibase 10.1103/PhysRevLett.119.100601} {\bibfield  {journal} {\bibinfo
  {journal} {Phys. Rev. Lett.}\ }\textbf {\bibinfo {volume} {119}},\ \bibinfo
  {pages} {100601} (\bibinfo {year} {2017})}\BibitemShut {NoStop}%
\bibitem [{\citenamefont {Rigol}\ \emph {et~al.}(2007)\citenamefont {Rigol},
  \citenamefont {Dunjko}, \citenamefont {Yurovsky},\ and\ \citenamefont
  {Olshanii}}]{Rigol2007}%
  \BibitemOpen
  \bibfield  {author} {\bibinfo {author} {\bibfnamefont {M.}~\bibnamefont
  {Rigol}}, \bibinfo {author} {\bibfnamefont {V.}~\bibnamefont {Dunjko}},
  \bibinfo {author} {\bibfnamefont {V.}~\bibnamefont {Yurovsky}}, \ and\
  \bibinfo {author} {\bibfnamefont {M.}~\bibnamefont {Olshanii}},\ }\href
  {\doibase 10.1103/PhysRevLett.98.050405} {\bibfield  {journal} {\bibinfo
  {journal} {Phys. Rev. Lett.}\ }\textbf {\bibinfo {volume} {98}},\ \bibinfo
  {pages} {050405} (\bibinfo {year} {2007})}\BibitemShut {NoStop}%
\bibitem [{\citenamefont {Cassidy}\ \emph {et~al.}(2011)\citenamefont
  {Cassidy}, \citenamefont {Clark},\ and\ \citenamefont {Rigol}}]{Cassidy2011}%
  \BibitemOpen
  \bibfield  {author} {\bibinfo {author} {\bibfnamefont {A.~C.}\ \bibnamefont
  {Cassidy}}, \bibinfo {author} {\bibfnamefont {C.~W.}\ \bibnamefont {Clark}},
  \ and\ \bibinfo {author} {\bibfnamefont {M.}~\bibnamefont {Rigol}},\ }\href
  {\doibase 10.1103/PhysRevLett.106.140405} {\bibfield  {journal} {\bibinfo
  {journal} {Phys. Rev. Lett.}\ }\textbf {\bibinfo {volume} {106}},\ \bibinfo
  {pages} {140405} (\bibinfo {year} {2011})}\BibitemShut {NoStop}%
\bibitem [{\citenamefont {Essler}\ and\ \citenamefont
  {Fagotti}(2016)}]{Essler2016}%
  \BibitemOpen
  \bibfield  {author} {\bibinfo {author} {\bibfnamefont {F.~H.~L.}\
  \bibnamefont {Essler}}\ and\ \bibinfo {author} {\bibfnamefont
  {M.}~\bibnamefont {Fagotti}},\ }\href
  {http://stacks.iop.org/1742-5468/2016/i=6/a=064002} {\bibfield  {journal}
  {\bibinfo  {journal} {J. Stat. Mech.}\ }\textbf {\bibinfo {volume} {2016}},\
  \bibinfo {pages} {064002} (\bibinfo {year} {2016})}\BibitemShut {NoStop}%
\bibitem [{Note1()}]{Note1}%
  \BibitemOpen
  \bibinfo {note} {We note that some studies have shown~\cite
  {Berman2004,Santos2012} that a kind of relaxation that is defined in terms of
  certain statistical indexes may not be qualitatively affected by
  integrability.}\BibitemShut {Stop}%
\bibitem [{\citenamefont {Barthel}\ and\ \citenamefont
  {Schollw\"ock}(2008)}]{Barthel2008}%
  \BibitemOpen
  \bibfield  {author} {\bibinfo {author} {\bibfnamefont {T.}~\bibnamefont
  {Barthel}}\ and\ \bibinfo {author} {\bibfnamefont {U.}~\bibnamefont
  {Schollw\"ock}},\ }\href {\doibase 10.1103/PhysRevLett.100.100601} {\bibfield
   {journal} {\bibinfo  {journal} {Phys. Rev. Lett.}\ }\textbf {\bibinfo
  {volume} {100}},\ \bibinfo {pages} {100601} (\bibinfo {year}
  {2008})}\BibitemShut {NoStop}%
\bibitem [{\citenamefont {Sotiriadis}\ \emph {et~al.}(2009)\citenamefont
  {Sotiriadis}, \citenamefont {Calabrese},\ and\ \citenamefont
  {Cardy}}]{Sotiriadis2009}%
  \BibitemOpen
  \bibfield  {author} {\bibinfo {author} {\bibfnamefont {S.}~\bibnamefont
  {Sotiriadis}}, \bibinfo {author} {\bibfnamefont {P.}~\bibnamefont
  {Calabrese}}, \ and\ \bibinfo {author} {\bibfnamefont {J.}~\bibnamefont
  {Cardy}},\ }\href {\doibase 10.1209/0295-5075/87/20002} {\bibfield  {journal}
  {\bibinfo  {journal} {{EPL} (Europhysics Letters)}\ }\textbf {\bibinfo
  {volume} {87}},\ \bibinfo {pages} {20002} (\bibinfo {year}
  {2009})}\BibitemShut {NoStop}%
\bibitem [{\citenamefont {Calabrese}\ \emph {et~al.}(2011)\citenamefont
  {Calabrese}, \citenamefont {Essler},\ and\ \citenamefont
  {Fagotti}}]{Calabrese2011}%
  \BibitemOpen
  \bibfield  {author} {\bibinfo {author} {\bibfnamefont {P.}~\bibnamefont
  {Calabrese}}, \bibinfo {author} {\bibfnamefont {F.~H.~L.}\ \bibnamefont
  {Essler}}, \ and\ \bibinfo {author} {\bibfnamefont {M.}~\bibnamefont
  {Fagotti}},\ }\href {\doibase 10.1103/PhysRevLett.106.227203} {\bibfield
  {journal} {\bibinfo  {journal} {Phys. Rev. Lett.}\ }\textbf {\bibinfo
  {volume} {106}},\ \bibinfo {pages} {227203} (\bibinfo {year}
  {2011})}\BibitemShut {NoStop}%
\bibitem [{\citenamefont {Calabrese}\ \emph
  {et~al.}(2012{\natexlab{a}})\citenamefont {Calabrese}, \citenamefont
  {Essler},\ and\ \citenamefont {Fagotti}}]{Calabrese2012_1}%
  \BibitemOpen
  \bibfield  {author} {\bibinfo {author} {\bibfnamefont {P.}~\bibnamefont
  {Calabrese}}, \bibinfo {author} {\bibfnamefont {F.~H.~L.}\ \bibnamefont
  {Essler}}, \ and\ \bibinfo {author} {\bibfnamefont {M.}~\bibnamefont
  {Fagotti}},\ }\href {\doibase 10.1088/1742-5468/2012/07/p07016} {\bibfield
  {journal} {\bibinfo  {journal} {Journal of Statistical Mechanics: Theory and
  Experiment}\ }\textbf {\bibinfo {volume} {2012}},\ \bibinfo {pages} {P07016}
  (\bibinfo {year} {2012}{\natexlab{a}})}\BibitemShut {NoStop}%
\bibitem [{\citenamefont {Calabrese}\ \emph
  {et~al.}(2012{\natexlab{b}})\citenamefont {Calabrese}, \citenamefont
  {Essler},\ and\ \citenamefont {Fagotti}}]{Calabrese2012_2}%
  \BibitemOpen
  \bibfield  {author} {\bibinfo {author} {\bibfnamefont {P.}~\bibnamefont
  {Calabrese}}, \bibinfo {author} {\bibfnamefont {F.~H.~L.}\ \bibnamefont
  {Essler}}, \ and\ \bibinfo {author} {\bibfnamefont {M.}~\bibnamefont
  {Fagotti}},\ }\href {\doibase 10.1088/1742-5468/2012/07/p07022} {\bibfield
  {journal} {\bibinfo  {journal} {Journal of Statistical Mechanics: Theory and
  Experiment}\ }\textbf {\bibinfo {volume} {2012}},\ \bibinfo {pages} {P07022}
  (\bibinfo {year} {2012}{\natexlab{b}})}\BibitemShut {NoStop}%
\bibitem [{\citenamefont {Cramer}\ \emph {et~al.}(2008)\citenamefont {Cramer},
  \citenamefont {Dawson}, \citenamefont {Eisert},\ and\ \citenamefont
  {Osborne}}]{Cramer2008}%
  \BibitemOpen
  \bibfield  {author} {\bibinfo {author} {\bibfnamefont {M.}~\bibnamefont
  {Cramer}}, \bibinfo {author} {\bibfnamefont {C.~M.}\ \bibnamefont {Dawson}},
  \bibinfo {author} {\bibfnamefont {J.}~\bibnamefont {Eisert}}, \ and\ \bibinfo
  {author} {\bibfnamefont {T.~J.}\ \bibnamefont {Osborne}},\ }\href {\doibase
  10.1103/PhysRevLett.100.030602} {\bibfield  {journal} {\bibinfo  {journal}
  {Phys. Rev. Lett.}\ }\textbf {\bibinfo {volume} {100}},\ \bibinfo {pages}
  {030602} (\bibinfo {year} {2008})}\BibitemShut {NoStop}%
\bibitem [{\citenamefont {Sotiriadis}\ and\ \citenamefont
  {Calabrese}(2014)}]{Sotiriadis2014}%
  \BibitemOpen
  \bibfield  {author} {\bibinfo {author} {\bibfnamefont {S.}~\bibnamefont
  {Sotiriadis}}\ and\ \bibinfo {author} {\bibfnamefont {P.}~\bibnamefont
  {Calabrese}},\ }\href {\doibase 10.1088/1742-5468/2014/07/P07024} {\bibfield
  {journal} {\bibinfo  {journal} {J. Stat. Mech.}\ }\textbf {\bibinfo {volume}
  {2014}},\ \bibinfo {pages} {P07024} (\bibinfo {year} {2014})}\BibitemShut
  {NoStop}%
\bibitem [{\citenamefont {Gluza}\ \emph {et~al.}(2016)\citenamefont {Gluza},
  \citenamefont {Krumnow}, \citenamefont {Friesdorf}, \citenamefont {Gogolin},\
  and\ \citenamefont {Eisert}}]{Gluza2016}%
  \BibitemOpen
  \bibfield  {author} {\bibinfo {author} {\bibfnamefont {M.}~\bibnamefont
  {Gluza}}, \bibinfo {author} {\bibfnamefont {C.}~\bibnamefont {Krumnow}},
  \bibinfo {author} {\bibfnamefont {M.}~\bibnamefont {Friesdorf}}, \bibinfo
  {author} {\bibfnamefont {C.}~\bibnamefont {Gogolin}}, \ and\ \bibinfo
  {author} {\bibfnamefont {J.}~\bibnamefont {Eisert}},\ }\href {\doibase
  10.1103/PhysRevLett.117.190602} {\bibfield  {journal} {\bibinfo  {journal}
  {Phys. Rev. Lett.}\ }\textbf {\bibinfo {volume} {117}},\ \bibinfo {pages}
  {190602} (\bibinfo {year} {2016})}\BibitemShut {NoStop}%
\bibitem [{\citenamefont {Bastianello}\ and\ \citenamefont
  {Sotiriadis}(2017)}]{Bastianello2017}%
  \BibitemOpen
  \bibfield  {author} {\bibinfo {author} {\bibfnamefont {A.}~\bibnamefont
  {Bastianello}}\ and\ \bibinfo {author} {\bibfnamefont {S.}~\bibnamefont
  {Sotiriadis}},\ }\href {\doibase 10.1088/1742-5468/aa5738} {\bibfield
  {journal} {\bibinfo  {journal} {J. Stat. Mech.}\ }\textbf {\bibinfo {volume}
  {2017}},\ \bibinfo {pages} {023105} (\bibinfo {year} {2017})}\BibitemShut
  {NoStop}%
\bibitem [{\citenamefont {Pozsgay}(2013)}]{Pozsgay2013}%
  \BibitemOpen
  \bibfield  {author} {\bibinfo {author} {\bibfnamefont {B.}~\bibnamefont
  {Pozsgay}},\ }\href {\doibase 10.1088/1742-5468/2013/07/P07003} {\bibfield
  {journal} {\bibinfo  {journal} {J. Stat. Mech.}\ }\textbf {\bibinfo {volume}
  {2013}},\ \bibinfo {pages} {P07003} (\bibinfo {year} {2013})}\BibitemShut
  {NoStop}%
\bibitem [{\citenamefont {Fagotti}\ and\ \citenamefont
  {Essler}(2013{\natexlab{a}})}]{Fagotti2013}%
  \BibitemOpen
  \bibfield  {author} {\bibinfo {author} {\bibfnamefont {M.}~\bibnamefont
  {Fagotti}}\ and\ \bibinfo {author} {\bibfnamefont {F.~H.}\ \bibnamefont
  {Essler}},\ }\href {\doibase 10.1088/1742-5468/2013/07/P07012} {\bibfield
  {journal} {\bibinfo  {journal} {J. Stat. Mech.}\ }\textbf {\bibinfo {volume}
  {2013}},\ \bibinfo {pages} {P07012} (\bibinfo {year}
  {2013}{\natexlab{a}})}\BibitemShut {NoStop}%
\bibitem [{\citenamefont {Fagotti}\ and\ \citenamefont
  {Essler}(2013{\natexlab{b}})}]{Fagotti-Essler2013}%
  \BibitemOpen
  \bibfield  {author} {\bibinfo {author} {\bibfnamefont {M.}~\bibnamefont
  {Fagotti}}\ and\ \bibinfo {author} {\bibfnamefont {F.~H.~L.}\ \bibnamefont
  {Essler}},\ }\href {\doibase 10.1103/PhysRevB.87.245107} {\bibfield
  {journal} {\bibinfo  {journal} {Phys. Rev. B}\ }\textbf {\bibinfo {volume}
  {87}},\ \bibinfo {pages} {245107} (\bibinfo {year}
  {2013}{\natexlab{b}})}\BibitemShut {NoStop}%
\bibitem [{\citenamefont {Goldstein}\ and\ \citenamefont
  {Andrei}(2014)}]{Goldstein2014}%
  \BibitemOpen
  \bibfield  {author} {\bibinfo {author} {\bibfnamefont {G.}~\bibnamefont
  {Goldstein}}\ and\ \bibinfo {author} {\bibfnamefont {N.}~\bibnamefont
  {Andrei}},\ }\href {\doibase 10.1103/PhysRevA.90.043625} {\bibfield
  {journal} {\bibinfo  {journal} {Phys. Rev. A}\ }\textbf {\bibinfo {volume}
  {90}},\ \bibinfo {pages} {043625} (\bibinfo {year} {2014})}\BibitemShut
  {NoStop}%
\bibitem [{\citenamefont {Pozsgay}(2014)}]{Pozsgay2014}%
  \BibitemOpen
  \bibfield  {author} {\bibinfo {author} {\bibfnamefont {B.}~\bibnamefont
  {Pozsgay}},\ }\href {\doibase 10.1088/1742-5468/2014/09/P09026} {\bibfield
  {journal} {\bibinfo  {journal} {J. Stat. Mech.}\ }\textbf {\bibinfo {volume}
  {2014}},\ \bibinfo {pages} {P09026} (\bibinfo {year} {2014})}\BibitemShut
  {NoStop}%
\bibitem [{\citenamefont {Wouters}\ \emph {et~al.}(2014)\citenamefont
  {Wouters}, \citenamefont {De~Nardis}, \citenamefont {Brockmann},
  \citenamefont {Fioretto}, \citenamefont {Rigol},\ and\ \citenamefont
  {Caux}}]{Wouters2014}%
  \BibitemOpen
  \bibfield  {author} {\bibinfo {author} {\bibfnamefont {B.}~\bibnamefont
  {Wouters}}, \bibinfo {author} {\bibfnamefont {J.}~\bibnamefont {De~Nardis}},
  \bibinfo {author} {\bibfnamefont {M.}~\bibnamefont {Brockmann}}, \bibinfo
  {author} {\bibfnamefont {D.}~\bibnamefont {Fioretto}}, \bibinfo {author}
  {\bibfnamefont {M.}~\bibnamefont {Rigol}}, \ and\ \bibinfo {author}
  {\bibfnamefont {J.-S.}\ \bibnamefont {Caux}},\ }\href {\doibase
  10.1103/PhysRevLett.113.117202} {\bibfield  {journal} {\bibinfo  {journal}
  {Phys. Rev. Lett.}\ }\textbf {\bibinfo {volume} {113}},\ \bibinfo {pages}
  {117202} (\bibinfo {year} {2014})}\BibitemShut {NoStop}%
\bibitem [{\citenamefont {Pozsgay}\ \emph {et~al.}(2014)\citenamefont
  {Pozsgay}, \citenamefont {Mesty\'an}, \citenamefont {Werner}, \citenamefont
  {Kormos}, \citenamefont {Zar\'and},\ and\ \citenamefont
  {Tak\'acs}}]{Pozsgay2014_correlations}%
  \BibitemOpen
  \bibfield  {author} {\bibinfo {author} {\bibfnamefont {B.}~\bibnamefont
  {Pozsgay}}, \bibinfo {author} {\bibfnamefont {M.}~\bibnamefont {Mesty\'an}},
  \bibinfo {author} {\bibfnamefont {M.~A.}\ \bibnamefont {Werner}}, \bibinfo
  {author} {\bibfnamefont {M.}~\bibnamefont {Kormos}}, \bibinfo {author}
  {\bibfnamefont {G.}~\bibnamefont {Zar\'and}}, \ and\ \bibinfo {author}
  {\bibfnamefont {G.}~\bibnamefont {Tak\'acs}},\ }\href {\doibase
  10.1103/PhysRevLett.113.117203} {\bibfield  {journal} {\bibinfo  {journal}
  {Phys. Rev. Lett.}\ }\textbf {\bibinfo {volume} {113}},\ \bibinfo {pages}
  {117203} (\bibinfo {year} {2014})}\BibitemShut {NoStop}%
\bibitem [{\citenamefont {Brockmann}\ \emph {et~al.}(2014)\citenamefont
  {Brockmann}, \citenamefont {Wouters}, \citenamefont {Fioretto}, \citenamefont
  {De~Nardis}, \citenamefont {Vlijm},\ and\ \citenamefont
  {Caux}}]{Brockmann2014}%
  \BibitemOpen
  \bibfield  {author} {\bibinfo {author} {\bibfnamefont {M.}~\bibnamefont
  {Brockmann}}, \bibinfo {author} {\bibfnamefont {B.}~\bibnamefont {Wouters}},
  \bibinfo {author} {\bibfnamefont {D.}~\bibnamefont {Fioretto}}, \bibinfo
  {author} {\bibfnamefont {J.}~\bibnamefont {De~Nardis}}, \bibinfo {author}
  {\bibfnamefont {R.}~\bibnamefont {Vlijm}}, \ and\ \bibinfo {author}
  {\bibfnamefont {J.-S.}\ \bibnamefont {Caux}},\ }\href {\doibase
  10.1088/1742-5468/2014/12/P12009} {\bibfield  {journal} {\bibinfo  {journal}
  {J. Stat. Mech.}\ }\textbf {\bibinfo {volume} {2014}},\ \bibinfo {pages}
  {P12009} (\bibinfo {year} {2014})}\BibitemShut {NoStop}%
\bibitem [{\citenamefont {Mesty{\'a}n}\ \emph {et~al.}(2015)\citenamefont
  {Mesty{\'a}n}, \citenamefont {Pozsgay}, \citenamefont {Tak{\'a}cs},\ and\
  \citenamefont {Werner}}]{Mestyan2015}%
  \BibitemOpen
  \bibfield  {author} {\bibinfo {author} {\bibfnamefont {M.}~\bibnamefont
  {Mesty{\'a}n}}, \bibinfo {author} {\bibfnamefont {B.}~\bibnamefont
  {Pozsgay}}, \bibinfo {author} {\bibfnamefont {G.}~\bibnamefont {Tak{\'a}cs}},
  \ and\ \bibinfo {author} {\bibfnamefont {M.}~\bibnamefont {Werner}},\ }\href
  {\doibase 10.1088/1742-5468/2015/04/P04001} {\bibfield  {journal} {\bibinfo
  {journal} {J. Stat. Mech.}\ }\textbf {\bibinfo {volume} {2015}},\ \bibinfo
  {pages} {P04001} (\bibinfo {year} {2015})}\BibitemShut {NoStop}%
\bibitem [{\citenamefont {Ilievski}\ \emph {et~al.}(2015)\citenamefont
  {Ilievski}, \citenamefont {De~Nardis}, \citenamefont {Wouters}, \citenamefont
  {Caux}, \citenamefont {Essler},\ and\ \citenamefont {Prosen}}]{Ilievski2015}%
  \BibitemOpen
  \bibfield  {author} {\bibinfo {author} {\bibfnamefont {E.}~\bibnamefont
  {Ilievski}}, \bibinfo {author} {\bibfnamefont {J.}~\bibnamefont {De~Nardis}},
  \bibinfo {author} {\bibfnamefont {B.}~\bibnamefont {Wouters}}, \bibinfo
  {author} {\bibfnamefont {J.-S.}\ \bibnamefont {Caux}}, \bibinfo {author}
  {\bibfnamefont {F.~H.~L.}\ \bibnamefont {Essler}}, \ and\ \bibinfo {author}
  {\bibfnamefont {T.}~\bibnamefont {Prosen}},\ }\href {\doibase
  10.1103/PhysRevLett.115.157201} {\bibfield  {journal} {\bibinfo  {journal}
  {Phys. Rev. Lett.}\ }\textbf {\bibinfo {volume} {115}},\ \bibinfo {pages}
  {157201} (\bibinfo {year} {2015})}\BibitemShut {NoStop}%
\bibitem [{\citenamefont {Murthy}\ and\ \citenamefont
  {Srednicki}(2018)}]{Murthy_arXiv2018}%
  \BibitemOpen
  \bibfield  {author} {\bibinfo {author} {\bibfnamefont {C.}~\bibnamefont
  {Murthy}}\ and\ \bibinfo {author} {\bibfnamefont {M.}~\bibnamefont
  {Srednicki}},\ }\href {https://arxiv.org/abs/1809.03681} {\bibfield
  {journal} {\bibinfo  {journal} {arXiv:1809.03681}\ } (\bibinfo {year}
  {2018})}\BibitemShut {NoStop}%
\bibitem [{\citenamefont {den Nijs}\ and\ \citenamefont
  {Rommelse}(1989)}]{Nijs1989}%
  \BibitemOpen
  \bibfield  {author} {\bibinfo {author} {\bibfnamefont {M.}~\bibnamefont {den
  Nijs}}\ and\ \bibinfo {author} {\bibfnamefont {K.}~\bibnamefont {Rommelse}},\
  }\href {\doibase 10.1103/PhysRevB.40.4709} {\bibfield  {journal} {\bibinfo
  {journal} {Phys. Rev. B}\ }\textbf {\bibinfo {volume} {40}},\ \bibinfo
  {pages} {4709} (\bibinfo {year} {1989})}\BibitemShut {NoStop}%
\bibitem [{\citenamefont {Hida}\ and\ \citenamefont {Takada}(1992)}]{Hida1992}%
  \BibitemOpen
  \bibfield  {author} {\bibinfo {author} {\bibfnamefont {K.}~\bibnamefont
  {Hida}}\ and\ \bibinfo {author} {\bibfnamefont {S.}~\bibnamefont {Takada}},\
  }\href {\doibase 10.1143/JPSJ.61.1879} {\bibfield  {journal} {\bibinfo
  {journal} {J. Phys. Soc. Jpn.}\ }\textbf {\bibinfo {volume} {61}},\ \bibinfo
  {pages} {1879} (\bibinfo {year} {1992})}\BibitemShut {NoStop}%
\bibitem [{\citenamefont {Hatsugai}(1992)}]{Hatsugai1992}%
  \BibitemOpen
  \bibfield  {author} {\bibinfo {author} {\bibfnamefont {Y.}~\bibnamefont
  {Hatsugai}},\ }\href {\doibase 10.1143/JPSJ.61.3856} {\bibfield  {journal}
  {\bibinfo  {journal} {J. Phys. Soc. Jpn.}\ }\textbf {\bibinfo {volume}
  {61}},\ \bibinfo {pages} {3856} (\bibinfo {year} {1992})}\BibitemShut
  {NoStop}%
\bibitem [{\citenamefont {Vidmar}\ and\ \citenamefont
  {Rigol}(2016)}]{Vidmar-Rigol2016}%
  \BibitemOpen
  \bibfield  {author} {\bibinfo {author} {\bibfnamefont {L.}~\bibnamefont
  {Vidmar}}\ and\ \bibinfo {author} {\bibfnamefont {M.}~\bibnamefont {Rigol}},\
  }\href {http://stacks.iop.org/1742-5468/2016/i=6/a=064007} {\bibfield
  {journal} {\bibinfo  {journal} {J. Stat. Mech.}\ }\textbf {\bibinfo {volume}
  {2016}},\ \bibinfo {pages} {064007} (\bibinfo {year} {2016})}\BibitemShut
  {NoStop}%
\bibitem [{Note2()}]{Note2}%
  \BibitemOpen
  \bibinfo {note} {The condition of subextensive fluctuations of macroscopic
  local conserved quantities is much weaker than the condition of the cluster
  decomposition property. The latter implies the former, but the former does
  not imply the latter.}\BibitemShut {Stop}%
\bibitem [{Note3()}]{Note3}%
  \BibitemOpen
  \bibinfo {note} {It should be noted that the converse is not true in
  general.}\BibitemShut {Stop}%
\bibitem [{Note4()}]{Note4}%
  \BibitemOpen
  \bibinfo {note} {Although Wick's theorem is usually applied to the vacuum
  state, it can be applied to individual eigenstates too in the present systems
  because the eigenstates can be expressed as a vacuum state by redefining the
  particles and holes for each mode. See Ref.~\cite
  {Molinari_arXiv2017}.}\BibitemShut {Stop}%
\bibitem [{\citenamefont {Molinari}(2017)}]{Molinari_arXiv2017}%
  \BibitemOpen
  \bibfield  {author} {\bibinfo {author} {\bibfnamefont {L.~G.}\ \bibnamefont
  {Molinari}},\ }\href {https://arxiv.org/abs/1710.09248} {\bibfield  {journal}
  {\bibinfo  {journal} {arXiv:1710.09248}\ } (\bibinfo {year}
  {2017})}\BibitemShut {NoStop}%
\bibitem [{Note5()}]{Note5}%
  \BibitemOpen
  \bibinfo {note} {The value $\protect \mathaccentV {tilde}003{\varepsilon
  }(p)+\protect \mathaccentV {tilde}003{\varepsilon }(-p)$ is non-negative for
  all $p$. In the case in which $\protect \mathaccentV {tilde}003{\varepsilon
  }(p)+\protect \mathaccentV {tilde}003{\varepsilon }(-p)$ touches the
  $p$-axis, the functions \protect \textup {\hbox {\mathsurround \z@ \protect
  \normalfont (\ignorespaces \ref {eq:fv}\unskip \@@italiccorr )}} and \protect
  \textup {\hbox {\mathsurround \z@ \protect \normalfont (\ignorespaces \ref
  {eq:fw}\unskip \@@italiccorr )}} may not be $L^1$-integrable. In this case,
  the following arguments remain valid by replacing $\protect \mathaccentV
  {tilde}003{\varepsilon }(p)+\protect \mathaccentV {tilde}003{\varepsilon
  }(-p)$ by a constant $\delta _c>0$ when $\protect \mathaccentV
  {tilde}003{\varepsilon }(p)+\protect \mathaccentV {tilde}003{\varepsilon
  }(-p)$ becomes smaller than $\delta _c$, and taking the limit $\delta _c\to
  0$ after $L\to \infty $ and $n_c\to \infty $}\BibitemShut {NoStop}%
\bibitem [{\citenamefont {Lazarides}\ \emph {et~al.}(2014)\citenamefont
  {Lazarides}, \citenamefont {Das},\ and\ \citenamefont
  {Moessner}}]{Lazarides2014a}%
  \BibitemOpen
  \bibfield  {author} {\bibinfo {author} {\bibfnamefont {A.}~\bibnamefont
  {Lazarides}}, \bibinfo {author} {\bibfnamefont {A.}~\bibnamefont {Das}}, \
  and\ \bibinfo {author} {\bibfnamefont {R.}~\bibnamefont {Moessner}},\ }\href
  {\doibase 10.1103/PhysRevLett.112.150401} {\bibfield  {journal} {\bibinfo
  {journal} {Phys. Rev. Lett.}\ }\textbf {\bibinfo {volume} {112}},\ \bibinfo
  {pages} {150401} (\bibinfo {year} {2014})}\BibitemShut {NoStop}%
\bibitem [{\citenamefont {Ishii}\ \emph {et~al.}(2018)\citenamefont {Ishii},
  \citenamefont {Kuwahara}, \citenamefont {Mori},\ and\ \citenamefont
  {Hatano}}]{Ishii-Kuwahara2018}%
  \BibitemOpen
  \bibfield  {author} {\bibinfo {author} {\bibfnamefont {T.}~\bibnamefont
  {Ishii}}, \bibinfo {author} {\bibfnamefont {T.}~\bibnamefont {Kuwahara}},
  \bibinfo {author} {\bibfnamefont {T.}~\bibnamefont {Mori}}, \ and\ \bibinfo
  {author} {\bibfnamefont {N.}~\bibnamefont {Hatano}},\ }\href {\doibase
  10.1103/PhysRevLett.120.220602} {\bibfield  {journal} {\bibinfo  {journal}
  {Phys. Rev. Lett.}\ }\textbf {\bibinfo {volume} {120}},\ \bibinfo {pages}
  {220602} (\bibinfo {year} {2018})}\BibitemShut {NoStop}%
\bibitem [{\citenamefont {D'Alessio}\ and\ \citenamefont
  {Rigol}(2014)}]{DAlessio2014}%
  \BibitemOpen
  \bibfield  {author} {\bibinfo {author} {\bibfnamefont {L.}~\bibnamefont
  {D'Alessio}}\ and\ \bibinfo {author} {\bibfnamefont {M.}~\bibnamefont
  {Rigol}},\ }\href {\doibase 10.1103/PhysRevX.4.041048} {\bibfield  {journal}
  {\bibinfo  {journal} {Phys. Rev. X}\ }\textbf {\bibinfo {volume} {4}},\
  \bibinfo {pages} {041048} (\bibinfo {year} {2014})}\BibitemShut {NoStop}%
\bibitem [{\citenamefont {Caux}\ and\ \citenamefont {Essler}(2013)}]{Caux2013}%
  \BibitemOpen
  \bibfield  {author} {\bibinfo {author} {\bibfnamefont {J.-S.}\ \bibnamefont
  {Caux}}\ and\ \bibinfo {author} {\bibfnamefont {F.~H.~L.}\ \bibnamefont
  {Essler}},\ }\href {\doibase 10.1103/PhysRevLett.110.257203} {\bibfield
  {journal} {\bibinfo  {journal} {Phys. Rev. Lett.}\ }\textbf {\bibinfo
  {volume} {110}},\ \bibinfo {pages} {257203} (\bibinfo {year}
  {2013})}\BibitemShut {NoStop}%
\bibitem [{\citenamefont {Ilievski}\ \emph {et~al.}(2016)\citenamefont
  {Ilievski}, \citenamefont {Quinn}, \citenamefont {De~Nardis},\ and\
  \citenamefont {Brockmann}}]{Ilievski2016}%
  \BibitemOpen
  \bibfield  {author} {\bibinfo {author} {\bibfnamefont {E.}~\bibnamefont
  {Ilievski}}, \bibinfo {author} {\bibfnamefont {E.}~\bibnamefont {Quinn}},
  \bibinfo {author} {\bibfnamefont {J.}~\bibnamefont {De~Nardis}}, \ and\
  \bibinfo {author} {\bibfnamefont {M.}~\bibnamefont {Brockmann}},\ }\href
  {\doibase 10.1088/1742-5468/2016/06/063101} {\bibfield  {journal} {\bibinfo
  {journal} {J. Stat. Mech.}\ }\textbf {\bibinfo {volume} {2016}},\ \bibinfo
  {pages} {063101} (\bibinfo {year} {2016})}\BibitemShut {NoStop}%
\bibitem [{\citenamefont {Berman}\ \emph {et~al.}(2004)\citenamefont {Berman},
  \citenamefont {Borgonovi}, \citenamefont {Izrailev},\ and\ \citenamefont
  {Smerzi}}]{Berman2004}%
  \BibitemOpen
  \bibfield  {author} {\bibinfo {author} {\bibfnamefont {G.~P.}\ \bibnamefont
  {Berman}}, \bibinfo {author} {\bibfnamefont {F.}~\bibnamefont {Borgonovi}},
  \bibinfo {author} {\bibfnamefont {F.~M.}\ \bibnamefont {Izrailev}}, \ and\
  \bibinfo {author} {\bibfnamefont {A.}~\bibnamefont {Smerzi}},\ }\href
  {\doibase 10.1103/PhysRevLett.92.030404} {\bibfield  {journal} {\bibinfo
  {journal} {Phys. Rev. Lett.}\ }\textbf {\bibinfo {volume} {92}},\ \bibinfo
  {pages} {030404} (\bibinfo {year} {2004})}\BibitemShut {NoStop}%
\bibitem [{\citenamefont {Santos}\ \emph {et~al.}(2012)\citenamefont {Santos},
  \citenamefont {Borgonovi},\ and\ \citenamefont {Izrailev}}]{Santos2012}%
  \BibitemOpen
  \bibfield  {author} {\bibinfo {author} {\bibfnamefont {L.~F.}\ \bibnamefont
  {Santos}}, \bibinfo {author} {\bibfnamefont {F.}~\bibnamefont {Borgonovi}}, \
  and\ \bibinfo {author} {\bibfnamefont {F.~M.}\ \bibnamefont {Izrailev}},\
  }\href {\doibase 10.1103/PhysRevLett.108.094102} {\bibfield  {journal}
  {\bibinfo  {journal} {Phys. Rev. Lett.}\ }\textbf {\bibinfo {volume} {108}},\
  \bibinfo {pages} {094102} (\bibinfo {year} {2012})}\BibitemShut {NoStop}%
\end{thebibliography}%

\end{document}